\documentclass[12pt]{iopart}
\usepackage{graphicx}
\usepackage{iopams}
\begin{document}

\title[Spatial coherence control and analysis via mixed-state ptychography]{Spatial coherence control and analysis via micromirror-based mixed-state ptychography}

\author{Ruslan R\"{o}hrich$^{1,2}$, A Femius Koenderink$^{2,*}$, Stefan M Witte$^{1,3}$ and Lars Loetgering$^{1,3}$}
\address{$^1$Advanced Research Center for Nanolithography, Science Park 106, 1098 XG Amsterdam, The Netherlands}
\address{$^2$Center for Nanophotonics, AMOLF, Science Park 104, 1098 XG Amsterdam, The Netherlands}
\address{$^3$Vrije Universiteit, De Boelelaan 1081, 1081 HV Amsterdam, The Netherlands}
\address{$^*$Author to whom any correspondence should be addressed.}
\ead{f.koenderink@amolf.nl}

\vspace{10pt}
\begin{indented} 
\item[]January 2020
\end{indented}

\begin{abstract}
Flexible and fast control of the phase and amplitude of coherent light, enabled by digital micromirror devices (DMDs) and spatial light modulators (SLMs), has been a driving force for recent advances in optical tweezers, nonlinear microscopy, and wavefront shaping. In contrast, engineering spatially partially coherent light remains widely elusive due to the lack of tools enabling a joint analysis and control sequence. Here, we report an approach to coherence engineering that combines a quasi-monochromatic, thermal source and a DMD together with a ptychographic scanning microscope. The reported method opens up new routes to low-cost coherence control, with applications in micromanipulation, nanophotonics, and quantitative phase contrast imaging.
\end{abstract}

\vspace{2pc}

%
\maketitle
%

\section{Introduction}
Optical coherence measures the degree of correlation that electromagnetic radiation exhibits in space and time~\cite{mandel1965coherence}. Theoretical progress combined with experimental control of partially coherent light continues to advance modern technology~\cite{Korotkova2020}. For instance, temporally partially coherent light is leveraged for depth-gating and three-dimensional imaging in optical coherence tomography, a major scientific and technological breakthrough of the last decades~\cite{Huang1178}. An example of the utility of partial spatial coherence is very long-baseline interferometry (VLBI), where measuring spatial correlations enables imaging far distance incoherent sources by means of the van Cittert-Zernike theorem~\cite{wolf2007introduction}. This has recently produced the first image of a black hole at radio frequencies~\cite{akiyama2019first}. These examples illustrate the importance of experimental access to both temporal and spatial coherence properties of electromagnetic radiation. However, for visible light the state of the art of coherence manipulation is arguably unbalanced. Measurement and control of temporally partially coherent light are possible thanks to the availability of high resolution spectroscopy, supercontinuum light sources, pulse shaping devices, and a plethora of nonlinear optical phenomena allowing for frequency conversion and broadening~\cite{weiner2011ultrafast}. The manipulation of spatially partially coherent light remains at a less advanced stage. While beam and coherence manipulating devices such as DMDs and SLMs are readily available, there are two primary factors that prevented efficient control of spatially partially coherent light. First, in contrast to VLBI, operating at radio frequencies and allowing to directly record phase, visible light coherence measurements require additional interferometric devices, which poses challenges in terms of stability. Second, a practical difficulty is the computational complexity of spatially partially coherent light, as it requires the specification of all two-point correlations within a beam cross-section, scaling with the fourth power of the number of samples per dimension upon discretization. This poses challenges both in terms of data acquisition and numerical processing.  

Among the experimental techniques to quantify spatial coherence, Young's double slit is most prominent~\cite{zernike1938concept,thompson1957two}. Several techniques based on this concept have been brought forward such as redundant~\cite{lin2003measurement} and non-redundant aperture arrays~\cite{mejia2007measuring} as well as non-parallel~\cite{divitt2014measuring} and programmable slits~\cite{partanen2014coherence, kondakci2017coherence}. However, despite significant improvements in speed, these techniques are either limited in resolution by the spacing of the apertures or slow due to their sequential acquisition scheme. Other methods such as lateral-shearing Sagnac interferometry (LSSI)~\cite{iaconis1996direct, naraghi2017wide} and phase-space tomography (PST)~\cite{raymer1994complex,waller2012phase,camara2013optical} require sophisticated optical elements, potentially introducing aberrations in the coherence function to be measured. An additional complication in LSSI is the long measurement time~\cite{iaconis1996direct}. PST is limited to paraxial optical fields and the experimental setup reported in~\cite{camara2013optical} required three SLMs, driving up cost and calibration effort. In addition, PST is computationally expensive as it seeks to recover a four-dimensional representation of a partially coherent beam.

In this work, we use ptychography~\cite{rodenburg2004phase}, a lensless imaging and wavefront sensing technique, for spatial coherence analysis. Ptychography has been demonstrated for label-free, quantitative phase imaging (QPI)~\cite{giewekemeyer2010quantitative, rose2018quantitative}, and beam characterization~\cite{guizar2008phase, thibault2008high, maiden2009improved,schropp2010hard,vila2014characterization, du2020measuring,loetgering2020generation}, with added benefits through various self-calibration methods~\cite{guizar2008phase, maiden2012annealing, zhang2013translation, loetgering2020zpie}.  An important extension is mixed-state ptychography, which uplifts the problem of fully coherent wavefront sensing to the computationally more complex problem of characterizing multiple incoherent contributions in a spatially partially coherent beam~\cite{thibault2013reconstructing}. This technique can be used to retrieve orthogonal modes in the decomposition of the mutual intensity (MI) of a spatially partially coherent beam~\cite{wolf1981new} and has become a standard tool in both x-ray and electron diffraction~\cite{enders2014ptychography, cao2016modal, chen2020mixed}. However, partial spatial coherence is considered a nuisance for high-resolution x-ray and electron diffraction imaging~\cite{enders2014ptychography,stachnik2015influence,li2018lensless,chen2020mixed}, despite flux advantages as compared to fully coherent radiation~\cite{li2016multiple}. In contrast, some visible light applications benefit from reduced spatial coherence. Examples include speckle noise reduction~\cite{redding2012speckle}, enhanced optical sectioning~\cite{tian20143d}, and partially coherent optical diffraction tomography~\cite{soto2017label}. A fast and low-cost experimental coherence control and analysis system could drive new insights in open research fields such as coherence transfer in nanophotonics~\cite{CHEN2020Optical} and optical tweezers with spatially partially coherent illumination~\cite{wang2007effect,aunon2013partially}. Here, we propose a combined spatial coherence analysis and control technique. Control is enabled by a LED-based DMD source, while analysis carried out via mixed state ptychography. Compared to PST, mixed state ptychography is computationally efficient, as it seeks to directly recover two-dimensional eigenfunctions of the mutual intensity, not the four-dimensional mutual intensity itself. We report, to our knowledge, both the first ptychography setup with controllable spatial coherence and the first quantitative comparison of experimental mixed-state ptychography to theoretical predictions. This enables us to fill an important gap in the literature, which is the verification of the reconstruction accuracy of mixed-state ptychography for spatially partially coherent fields.

\section{Methods}
\subsection{Spatial coherence theory\label{sec:theory}}
Throughout this paper, we restrict the discussion to quasi-monochromatic light, which allows us to omit the frequency dependence of all fields involved. Under this assumption, spatial coherence is quantified using the mutual intensity ${J}(\mathbf{r}_1,\mathbf{r}_2)$, which captures correlations between pairs of points in two independent spatial variables $\mathbf{r}_1$ and $\mathbf{r}_2$. These spatial correlations can be decomposed into an orthonormal basis  $\left\{ \phi_{m}\right\}$ with $m=1,...,M$,
\begin{equation}
{J}(\mathbf{r}_1,\mathbf{r}_2) = {\displaystyle \sum_{m} \lambda_m \phi_m^{*}(\mathbf{r}_1) \phi_m(\mathbf{r}_2) },
\label{eq:mutualint}
\end{equation}
where the `modes' $\phi_{m}$ are eigenfunctions with eigenvalues $\lambda_m$ as reviewed in Refs.~\cite{martinez1979expansion,wolf1981new}. The normalized MI  
\begin{equation}
\gamma(\mathbf{r}_1,\mathbf{r}_2) = \frac{{J}(\mathbf{r}_1,\mathbf{r}_2) }{\sqrt{I(\mathbf{r}_1)I(\mathbf{r}_2)}}
\label{eq:norm_mutualint}
\end{equation}
is bounded in magnitude by 0 and 1 and will be used in the following as a measure for spatial coherence. The distribution of the eigenvalues $\lambda_m$, which expresses the occupancy of the different modes, can be used to calculate an alternative coherence metric given by
\begin{equation}
 \bar{\mu}^2 =  \frac{ {\sum_{m} \lambda_m^2}  }{ {\left( \sum_{m} \lambda_m \right)^2} }.
\label{eq:mu}
\end{equation}
This quantity $\bar{\mu} \in [0, 1]$ is referred to as the overall coherence.  The inverse square of $\bar{\mu}$ provides an upper bound for the number of coherent modes required for an accurate mode representation in Eq.~\eref{eq:mutualint}~\cite{starikov1982effective}. In the following, we refer to this quantity as the effective number of modes $M_{\mathrm{eff}}= 1/\bar{\mu}^2$. 

\subsection{Mixed-state ptychography}
Ptychography is based on measurements of diffraction patterns generated by the interaction of a loosely focused probe beam $P\left(\boldsymbol{r}\right)$ and a scattering object $O\left(\boldsymbol{r}\right)$, which is laterally scanned while a series of diffraction patterns is recorded~\cite{rodenburg2004phase}. The scan step is chosen small enough to guarantee adjacent diffraction signatures are correlated. High overlap promotes overdetermination in the inverse problem underlying ptychography, allowing to solve the inverse scattering problem by iteratively recovering the complex probe and object transmission functions~\cite{thibault2008high, maiden2009improved}. The mixed-state formulation of ptychography expands on this concept by recovering the orthogonal modes $P_{m}\left(\boldsymbol{r}\right)=\sqrt{\lambda_{m}}\phi_{m}\left(\boldsymbol{r}\right)$ in the expansion of the MI of a spatially partially coherent illumination~\cite{thibault2013reconstructing}. Partial spatial coherence manifests itself as reduced visibility in the diffraction intensities observed in ptychography. However, the loss of contrast in diffraction patterns is not exclusively caused by spatial decoherence but may for instance be caused by object vibrations or temporal incoherence~\cite{enders2014ptychography,clark2014dynamic}.  
Therefore, in order to single out spatial incoherence one has to take some precautions by e.g., ensuring object stability and by spectrally filtering the light source. The forward model of ptychography assumes the exit wave downstream of the specimen ($\boldsymbol{r}$) as product of the estimated probe and the object functions 
\begin{equation}
\psi^{n}_{m,j}(\mathbf{r})=P_{m}^n(\mathbf{r})O^{n}(\mathbf{r} -\mathbf{t_j}),
  \label{eq:psi}
\end{equation}
where $n$ is the iteration index and $\boldsymbol{t}_{j}$ is a translation vector. The exit waves for each illumination mode are propagated into the detector plane ($\boldsymbol{q}$) and incoherently added 
\begin{equation}
\label{eq:detectorIncoherentSum}
    I\left(\boldsymbol{q}\right)=\sum_{m}\left|\tilde{\psi}_{m}\left(\boldsymbol{q}\right)\right|^{2},
\end{equation}
where $\psi_{m}\left(\boldsymbol{r}\right)$ and $\tilde{\psi}_{m}\left(\boldsymbol{q}\right)$ are related through free-space propagation. The forward model is inverted by iteratively applying gradient steps in the detector and object planes~\cite{maiden2009improved, thibault2013reconstructing}, to wit
\begin{equation}
    \tilde{\psi}_{m,j}^{n+1}=\frac{\sqrt{I_j}}{\sqrt{\sum_{m'}\left|\tilde{\psi}_{m',j}^{n}\right|^{2}}}\tilde{\psi}_{m,j}^{n},
\end{equation}
\begin{equation}
    P_{m}^{n+1}=P_{m}^{n}+\frac{\beta}{\max\left|O^{n}\right|^{2}}O^{n*}\left(\psi_{m,j}^{n+1}-P_{m}^{n}O^{n}\right),
    \label{eq:Pupdate}
\end{equation}

\begin{equation}
    O^{n+1}=O^{n}+\frac{\beta}{\max\sum_{m'}\left|P_{m'}^{n}\right|^{2}}\sum_{m}P_{m}^{n*}\left(\psi_{m,j}^{n+1}-P_{m}^{n}O^{n}\right),
    \label{eq:Oupdate}
\end{equation}
where $\beta$ is a parameter that controls the step size of the probe and object updates. We omitted the functional dependencies for notational simplicity. The reconstructed probe state mixtures are iteratively orthogonalized, for instance using the Gram-Schmidt procedure or singular value decomposition. Further algorithmic details are provided in Refs.~\cite{thibault2013reconstructing, loetgering2017data}.

\subsection{Experimental setup}
\begin{figure}[p]
\centering
\includegraphics[width=0.66\linewidth]{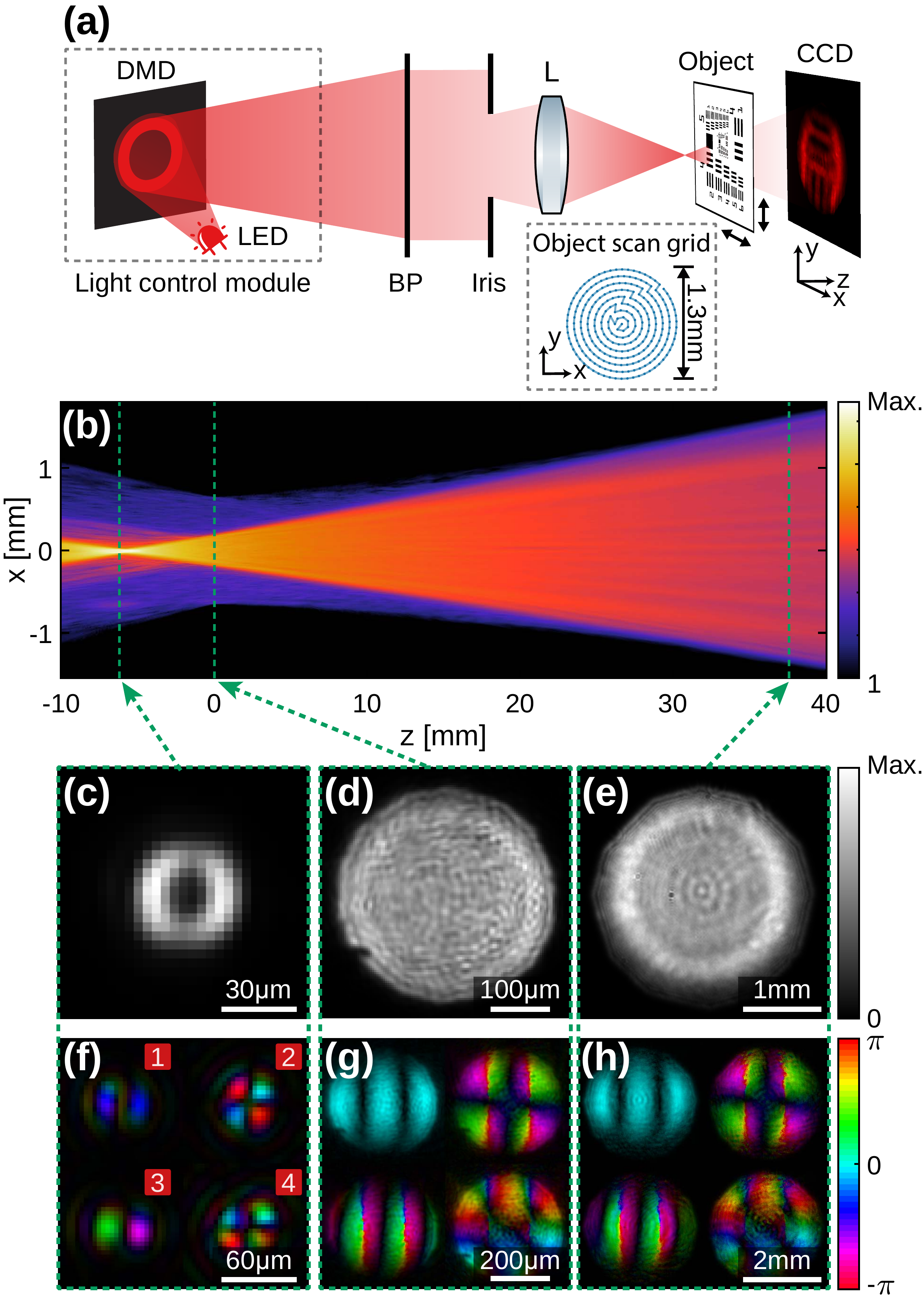}
\caption{(a) Sketch of the micromirror-based mixed-state ptychography setup. The light control module consisting of an LED source and a DMD generates a beam with a binary amplitude pattern. This beam is spectrally filtered using a bandpass (BP) centered at $\lambda=632.8\,$nm ($1\,$nm FWHM) and defined by an iris. A lens (L) images the DMD upstream of an object, which is mounted on a translation stage. The inset indicates the concentric object scan grid, which consists of 260 points. A sequence of diffraction intensities is recorded on a CCD. (b) The reconstructed probe intensity in the xz plane. The planes of the DMD image, object and CCD are indicated as dashed lines. The intensity is shown on a logarithmic color scale. (c)-(d) The probe intensities in the xy planes of the DMD, object and CCD, respectively. (f)-(h) The first four out of a total of 36 reconstructed coherent modes in the planes of (c)-(d). The amplitude is encoded as brightness and phase as hue. The phase curvature of the probe modes in panels (g) and (h) was removed. In panel (a) for simplicity the reflective DMD is shown operating in transmission.}
\label{fig:setup}
\end{figure}

The setup used for joint spatial coherence control and analysis is depicted in figure~\ref{fig:setup}(a). A light control module (Texas Instruments DLP3010EVM-LC) combining an RGB LED source and a DMD is used to generate a beam with static, binary amplitude pattern encoded into it. The DMD consists of an array of $1280\times720$ micromirrors with a pitch of $5.4$\,\textmu m.
The micromirror pixels of the DMD are in the following denoted as dpx. We switch on only the red LED and further spectrally filter the light using a bandpass (BP) filter centered at $\lambda=632.8\,$nm (FWHM of $1\,$nm) further downstream. An iris is used to limit the beam size to approximately half the angular range of the detector. After this a lens ($100\,$mm focal length) is focusing the probe beam, forming an image of the DMD $6.2\,$mm upstream of the object plane. This image is demagnified by approximately a factor 2 as compared to the original pattern displayed on the DMD. 

The object, a USAF resolution test target (Thorlabs R3L1S4P), is mounted on an xy-translation stage (Smaract SLC-1770-D-S, $70\,$nm repeatability). Finally, the scattered signal at each scan position is recorded by a CCD camera (AVT prosilica GX1920, $14\,$bit, $4.54\,$\textmu m pixel size). The ptychography scans were performed on a concentric grid (shown as an inset in figure~\ref{fig:setup}(a)). The object scan grid consists of $260$ steps with a step size of $70$\,\textmu m and a FOV of $1.3\times1.3$\,mm$^2$. Since in all our measurements the minimally observed beam size in the object plane had an FWHM of $262$\,\textmu m, this resulted in a linear probe overlap of at least $73\,\%$. Further setup implementation details are given in~\ref{sec:setupdetails}.

\section{Results}
\subsection{Setup characterization}
The coherent mode representation offers flexibility in the characterization of partially coherent optical systems, as it allows to numerically propagate the beam to any desired propagation distance once a ptychography reconstruction has been completed. For propagation we use the band-limited angular spectrum method described in Ref.~\cite{matsushima2009band}. The observable intensity at a particular plane is calculated by individually propagating the coherent modes  before summing them incoherently, as described by Eq.~\eref{eq:detectorIncoherentSum}. In figure~\ref{fig:setup}(b) we show such a propagation result by plotting an xz-cross-section the intensity on a logarithmic scale. The plot displays a $z$ range of $-10\,$mm to $40\,$mm with a step size of $0.1$\,mm, whereby $z=0$ is defined as the object plane. In this example, we use a ptychographic reconstruction of a measurement using a DMD image of a ring as its partially coherent source. The exact DMD image displayed a ring with an outer diameter of $15$\,dpx and width of $2$\,dpx, where dpx denotes DMD pixel size. Figures~\ref{fig:setup}(c)-\ref{fig:setup}(e) show the reconstructed probe intensities at three z positions of particular interest. The corresponding first four coherent modes are presented in figures~\ref{fig:setup}(f)-\ref{fig:setup}(h). In figure~\ref{fig:S4_modes}(a-c) we show all 36 reconstructed coherent modes of this example, where one can see that higher order modes contain the noisy part of the signal. Figure~\ref{fig:setup}(c) displays the reconstructed intensity at $z=-6.2\,$mm, which is an image plane of the DMD. The $z$ position is indicated in figure~\ref{fig:setup}(b) by a vertical dashed line.  In addition to the mostly focused beam, the logarithmic color plot in figure~\ref{fig:setup}(b) reveals the presence of background light, which is caused by the imperfect amplitude contrast of the DMD. This effect, which becomes more prominent when fewer DMD pixels are activated, did not hinder successful reconstructions since the background signal is usually captured by higher order coherent modes. The $z$ coordinate of the DMD image plane was found manually by comparing the sharpness of the DMD pattern reconstruction in steps of $0.1$\,mm. The demagnification factor of around $0.5$ was obtained by comparing the size of the reconstructed DMD image to the size of the programmed DMD image (ring diameter multiplied by the DMD pixel size of $5.4\,$\textmu m). Figure~\ref{fig:setup}(d) shows the probe intensity at the object plane ($z=0$), which for instance can be used to estimate the average probe overlap. The probe intensity in the CCD plane ($z=37.6\,$mm) is shown in figure~\ref{fig:setup}(e). The probe was slightly cropped by an iris, which possesses an image plane further downstream at around $z\approx50\,$mm. The sample-detector distance was calibrated using the method described in Ref.~\cite{loetgering2020zpie}. To aid visual inspection of the coherent modes in figures~\ref{fig:setup}(g) and \ref{fig:setup}(h), the quadratic phase curvature was removed through multiplication with the conjugate phase of the primary coherent mode.

\begin{figure}[t!]
\centering
\includegraphics[width=0.68\linewidth]{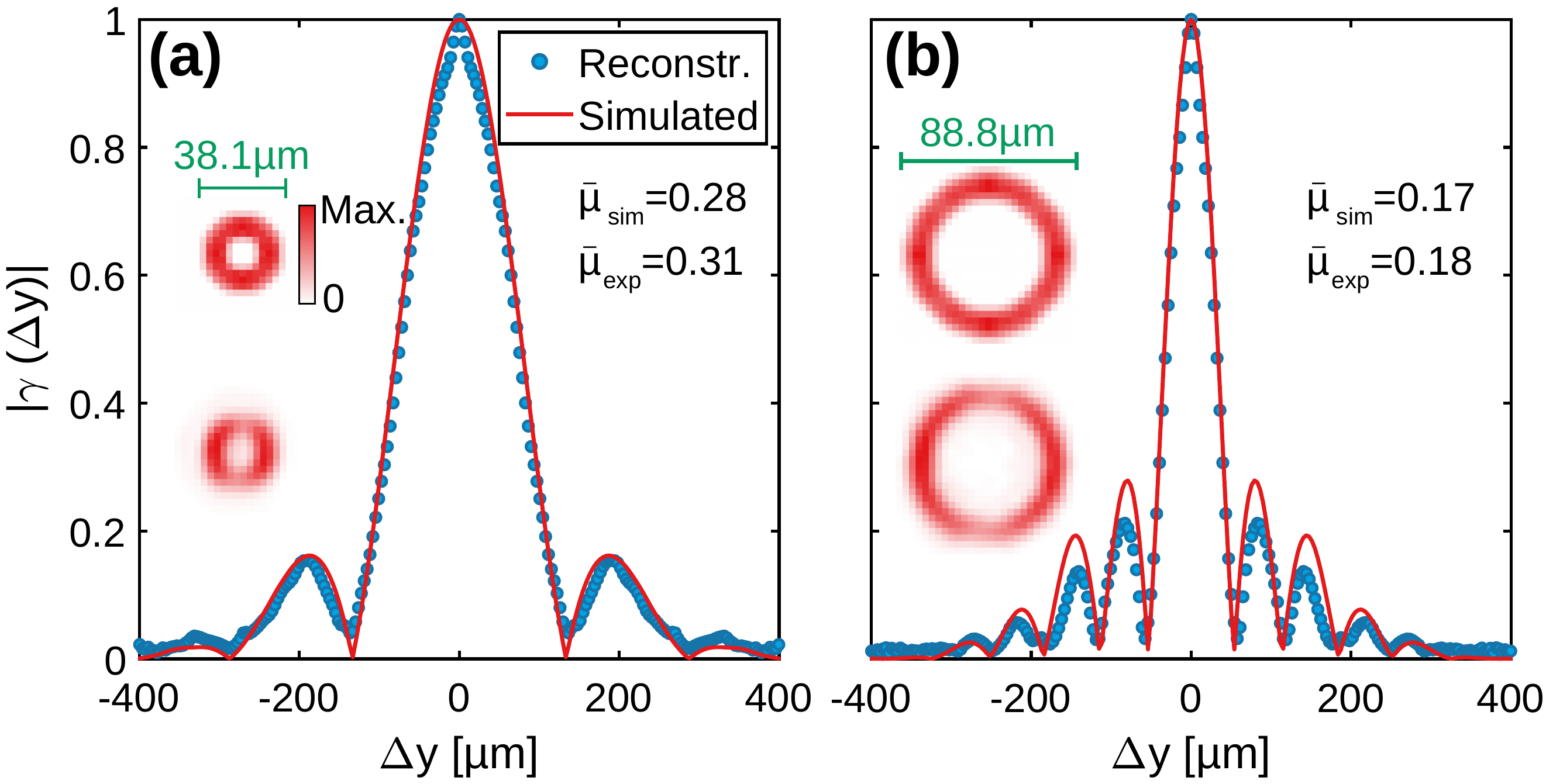}
\caption{Measured and simulated magnitude of the normalized MI $|\gamma(\Delta y)|$ of two differently sized ring-shaped sources at $z=10$\,mm. The ring diameters in the DMD images are $15$\,dpx (a) and $35$\,dpx (b). The insets show the simulated (top) and reconstructed (bottom) source intensity distributions together as well as their overall coherence $\bar{\mu}$.}
\label{fig:ringSimulation}
\end{figure}

\subsection{Quantitative coherence analysis}
In this section, we first experimentally examine the coherence analysis capabilities of mixed-state ptychography by comparison to simulation results. Following this, we demonstrate how the two metrics described in Sec.~2\ref{sec:theory}, the MI $\gamma(\Delta y)$ and the overall coherence $\bar{\mu},$ can be controlled by variation of the source shape using the DMD. 

In figure~\ref{fig:ringSimulation} we compare experimentally obtained curves for $|\gamma(\Delta y)|$ (blue dots) to simulations (red line). The DMD images, used in these measurements are composed of a ring with a thickness of $2\,$dpx and the outer diameters $d=15$\,dpx and $d=35$\,dpx. We model these two extended sources as a collection of point sources, that emit mutually incoherent spherical waves from a ring-shaped area at $z=-6.2\,$mm (conjugate plane of the DMD). To model the intensity distribution in this source area, we first generate a binary amplitude mask and then blur its edges using a circular averaging filter with a radius of $2$\,px, approximating the low-pass filtering operation of the iris upstream of the lens in figure~\ref{fig:setup}(a). Here px denotes the probe pixel size, which in both the simulated and reconstructed source images is $3.6\,$\textmu m. The resulting simulated source intensities are shown in the insets of figure~\ref{fig:ringSimulation} above the corresponding experimentally reconstructed source intensities. All non-zero pixels of the simulated DMD plane act as mutually incoherent point sources. At the iris plane ($z=50\,$mm) a circular mask with a diameter of $3.6\,$mm is applied. Performing a singular value decomposition (SVD) on the set of spherical waves, an orthogonal mode representation of the spatially partially coherent beam is found. Finally, we propagate both the simulated and measured probe modes to $z=10$\,mm and calculate the 2D normalized MI $\gamma(\mathbf{y}_1,\mathbf{y}_2)$ as given by Eq.~\eref{eq:norm_mutualint} at $x=0$. To reduce the influence of noise on the 1D cross-sections $\gamma(\Delta y)$ shown in figure~\ref{fig:ringSimulation} we average over the first 18 pixels above and below the main diagonal of the 2D mutual coherence. The simulated and measured magnitude of the normalized MI in figure~\ref{fig:ringSimulation} show an excellent agreement, quantitatively verifying the coherence measurement abilities of mixed-state ptychography.



\begin{figure*}[ht]
\centering
\includegraphics[width=1\linewidth]{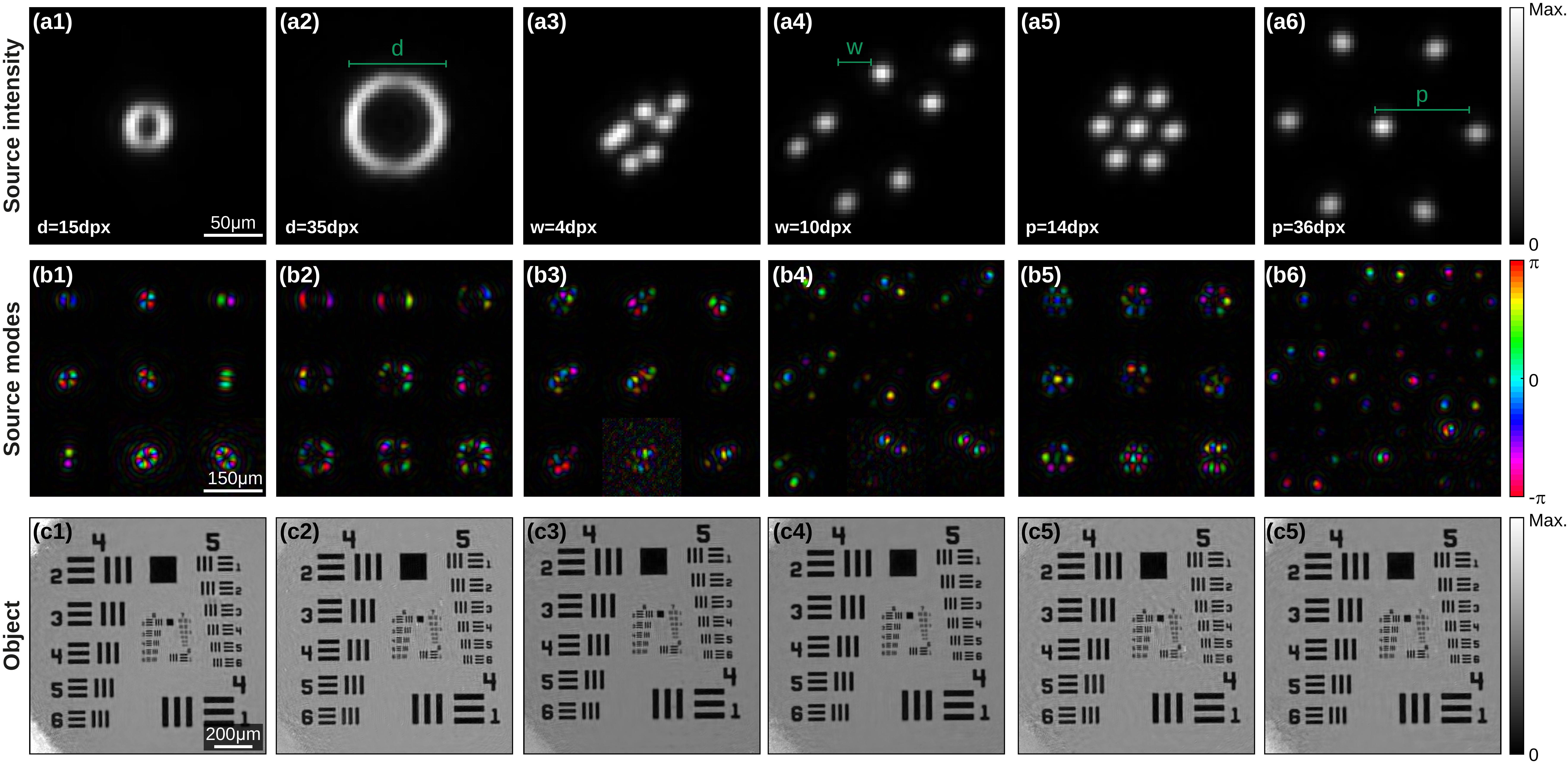}
\caption{Reconstruction results for ring, Costas and hexagonal array sources. Intensity distribution (a1)-(a6) and first 9 coherent modes (b1)-(b6) at the DMD plane. (c1)-(c6) Reconstructed object intensity.\label{fig:sourceGeometry}}
\end{figure*}

We performed additional measurements to investigate the influence of the source shape on the MI and its associated coherence measures. In addition to the ring shape, we chose two source shapes to which we refer as Costas~\cite{costas1984study} and hexagonal arrays. Both the Costas and the hexagonal array consist of 7 square elements but in different spatial arrangements. Costas arrays are latin squares in which every displacement vector is unique. Hexagonal arrays on the other hand consist of square elements arranged at the vertices and center of a regular hexagon. To give an overview over the investigated source shapes, 6 examples of reconstructed probe intensities in the DMD plane are shown in figures~\ref{fig:sourceGeometry}(a1)-\ref{fig:sourceGeometry}(a6), where each source shape is present in two varying choices of size parameters ($d,w,p$). Note, that the successful reconstruction of the different source shapes through back-propagation of coherent modes further supports the validity of the presented results. The corresponding first $9$ coherent modes in the DMD plane are shown in figures~\ref{fig:sourceGeometry}(b1)-\ref{fig:sourceGeometry}(b6). These modes can be used to gain insight into spatial correlations in the partially coherent source. Our data shows such insight beyond the fully expected localization of the reconstructed modes on the DMD mask, as an unexpected symmetry breaking, whereby the first modes preferentially show zero crossings in the horizontal direction. We attribute this to ensemble decoherence effects as a result of switching between on and off state of the micromirrors during the pattern streaming cycle, which appears even in static mode for the DMD control module deployed in our setup.


\begin{figure*}[ht]
\centering
\includegraphics[width=1\linewidth]{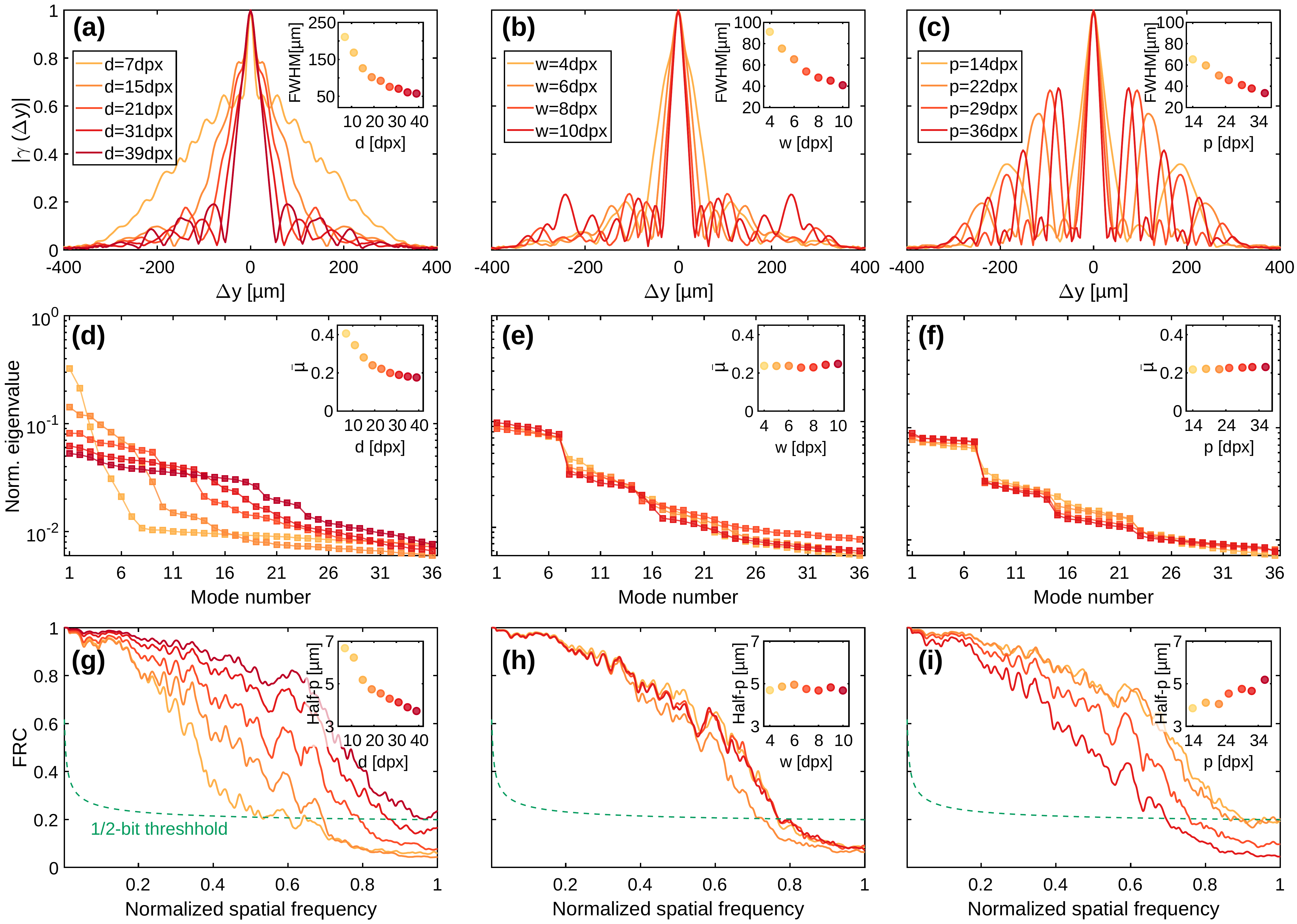}
\caption{Quantitative characterization of reconstruction results for ring, Costas and hexagonal array sources. (a)-(c) Magnitude of the normalized MI at $z=10$\,mm. The insets show FWHM of $|\gamma(\Delta y)|$ as a function of the respective scaling parameter. (d)-(f) Normalized eigenvalue distribution plotted on a logarithmic scale. The inset panels show the overall coherence $\bar{\mu}$. (g)-(i) Object Fourier ring correlation (FRC) as a function of normalized spatial frequency. The half-period resolution at the intersection of the FRC and the $1/2$-bit curves is shown in inset plots.\label{fig:coherence_charact}}
\end{figure*}

In figure~\ref{fig:coherence_charact}(a)-\ref{fig:coherence_charact}(c) we show reconstructed MI cross-sections for various sizes of the ring, Costas and hexagonal array sources, evaluated at $z=10$\,mm. Supporting details on DMD control are given in~\ref{sec:DMD}. The results in figures~\ref{fig:coherence_charact}(a) to \ref{fig:coherence_charact}(c) demonstrate our ability to control and measure the MI $|\gamma(\Delta y)|$. From the insets we observe that increasing the effective source size leads to a decreasing FWHM of $|\gamma(\Delta y)|$ and therefore lower spatial coherence lengths. Since the sizes of the hexagonal arrays were chosen such that the largest separations approximately matched the ones in the Costas array, they possess FWHM values of a similar range. The 1D MI cross-section of the hexagonal array features side lobes, which are less pronounced in the other two source shapes. This can be understood as a consequence of the van-Cittert-Zernike theorem, which implies a Fourier-transform relationship between an incoherent source and its far-field mutual intensity~\cite{wolf2007introduction}. Thus, periodic sources exhibit more pronounced correlation peaks than aperiodic sources, as seen here by comparing the MI of the hexagonal array with the MI of the Costas array.

The semi-logarithmic plots in figures~\ref{fig:coherence_charact}(d) to \ref{fig:coherence_charact}(f) show the distribution of the eigenvalues $\lambda_m$ (compare Eq.~\eref{eq:mutualint}) over the first $m=1,...,36$ modes for the three source shapes. The sum over the $\lambda_m$ values was normalized to unity. For the ring-shaped source in figure~\ref{fig:coherence_charact}(d), a higher diameter $d$ continuously increases the number of occupied modes, due to the larger number of source points contained in the source. For the Costas and hexagon array sources in figures~\ref{fig:coherence_charact}(e) and \ref{fig:coherence_charact}(f) the mode occupancy remains nearly unchanged during the size sweep, which is due to a fixed number of source points. In addition, we observe that the eigenvalues of the Costas and hexagonal array undergo step-like changes between several plateaus of approximately constant values at integer-multiples of 7, indicating partial spatial coherence over each of the 7 active sub-areas displayed on the DMD. The insets in figures~\ref{fig:coherence_charact}(d) to \ref{fig:coherence_charact}(f) show that the overall coherence for the Costas and hexagonal arrays are approximately equal. For the ring-shaped source the overall coherence decreases from $\bar{\mu}=0.41$ to $\bar{\mu}=0.18$ as a function of $d$, which corresponds to an increase in effective modes from $M_{\mathrm{eff}}=6$ to $M_{\mathrm{eff}}=32$. In case of the Costas array the overall coherence remains nearly constant and fluctuates around $\bar{\mu}=0.24$, which corresponds to $M_{\mathrm{eff}}= 17$. For the hexagonal array the overall coherence fluctuates around $\bar{\mu}=0.22$, which corresponds to $M_{\mathrm{eff}}= 20$.

\subsection{Influence of illumination on object reconstruction}
Above we have focused on the characterization of partially coherent illumination. In this section, we examine several aspects regarding the influence of partial spatial coherence on the object reconstruction quality. Structured illumination has previously been reported to offer advantages for fully coherent ptychography, due to relaxed detector dynamic range requirements~\cite{maiden2010optical, maiden2011superresolution,guizar2012role,li2016multiple,odstrvcil2019towards}. First, we investigate whether controlling the structure of partially coherent illumination affects the reconstruction quality. Second, by varying the source size we effectively alter degrees of coherence of the illumination. Previous studies have suggested that decreased spatial coherence generally leads to reduced reconstruction quality~\cite{stachnik2015influence}. We test how the lateral spatial resolution depends on the illumination spatial coherence under variation of the source shape. Third, we compare the reproducibility of QPI via ptychography, by comparing reconstruction obtained through fully coherent and mixed-state ptychography.

We use Fourier ring correlation (FRC) as a function of spatial frequency to quantify resolution~\cite{van2005fourier}. In this method, two object reconstructions, which are independently acquired under the same experimental conditions, are correlated on concentric rings in Fourier space. Here, we obtain these two independent object reconstructions by splitting the data into two half sets, each containing 130 diffraction patterns (see~\ref{sec:recdetails}). Figures~\ref{fig:coherence_charact}(g)-\ref{fig:coherence_charact}(i) summarize the FRC results for the same three source shapes as discussed in the last section with various sizes. The horizontal axis shows the spatial frequency as a fraction of the Nyquist limit. The spatial frequency at the intersection between the FRC and the 1/2-bit information threshold curve is converted into a half-period resolution value and used as a measure for spatial resolution~\cite{van2005fourier}. The insets of figures~\ref{fig:coherence_charact}(g)-\ref{fig:coherence_charact}(i) show all half-period results as a function of each respective scaling parameter. The results with a ring source in figure~\ref{fig:coherence_charact}(g) show an improved FRC for increasing ring diameters $d$, with the corresponding half-period resolution shrinking from $6.7\,$\textmu m to $d=3.7\,$\textmu m. The reason for this behavior is the increased photon flux as a function of ring diameter, see figure~\ref{fig:S1_flux}(c). Two examples of reconstructed objects with a small and large ring diameter are shown in figure~\ref{fig:sourceGeometry}(c1) and \ref{fig:sourceGeometry}(c2), respectively.  Next, in figures~\ref{fig:coherence_charact}(h) and \ref{fig:coherence_charact}(i) we show FRC results for the Costas and the hexagonal source arrays. Here the source size variation changes the angles of illumination, but keeps the total source area constant. The inset of figure~\ref{fig:coherence_charact}(h) shows the half-period resolution extracted from the FRC results with Costas array illumination. The values do not show a clear trend but rather fluctuate around a resolution of $4.8\,$\textmu m. For the hexagonal array the half-period resolution shown in the inset  of figure~\ref{fig:coherence_charact}(h) seems to increase from $3.8\,$\textmu m to $5.2\,$\textmu m  as a function of $p$. This is due to a decrease of incoming flux with larger $p$ values as shown in figure~\ref{fig:S1_flux}(i), caused by apodization of the source by the iris upstream of the lens in figure~\ref{fig:setup}(a).

\begin{figure}[t!]
\centering
\includegraphics[width=0.9\linewidth]{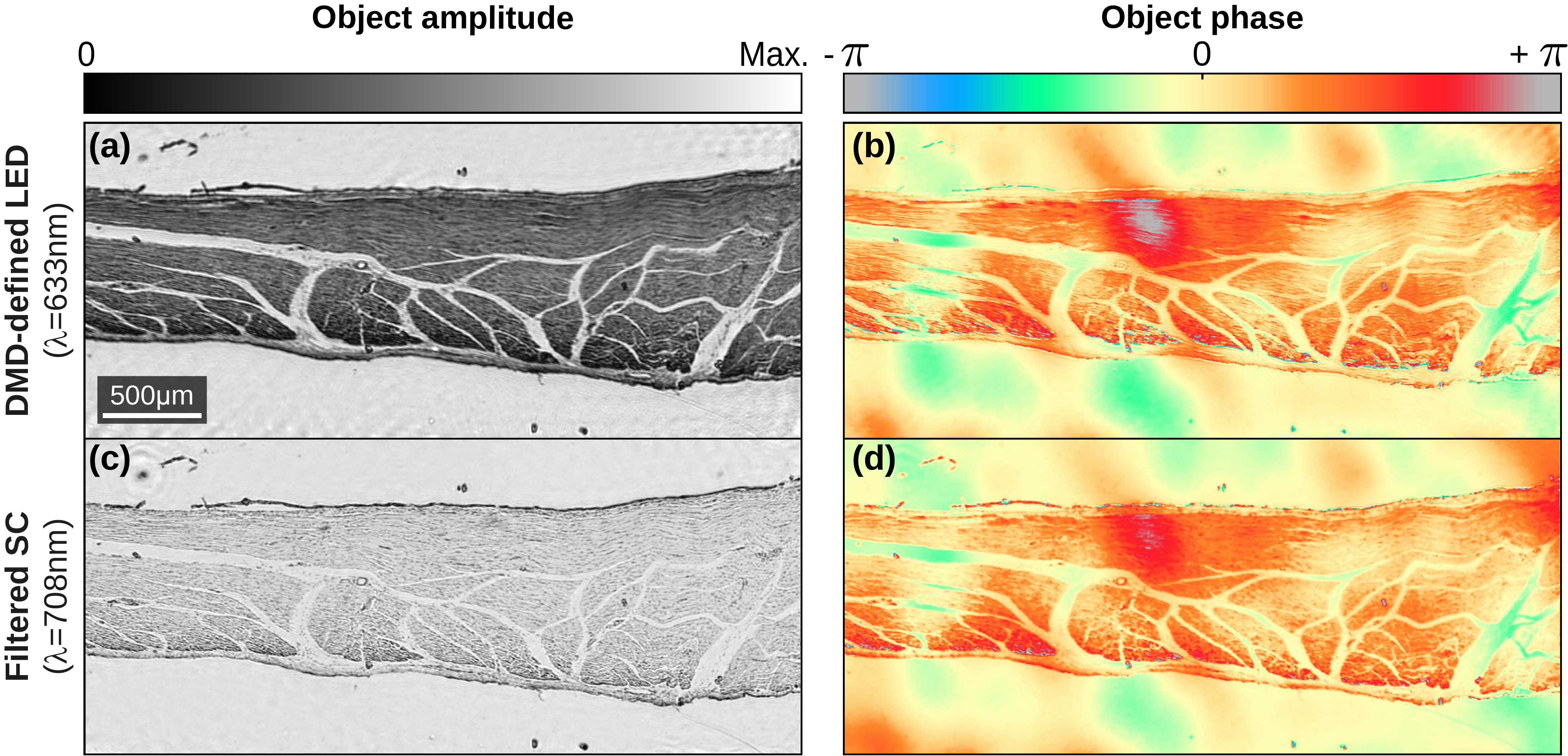}
\caption{Amplitude and phase contrast of a histological section of a human muscle imaged with mixed-state ptychography (a) and (b) as well as using fully coherent ptychography (c) and (d).}
\label{fig:muscle}
\end{figure}

Finally, we replaced the USAF target with a biological sample containing a histological section of a human muscle. Since this sample features richer phase variations compared to the USAF target (binary amplitude object), it is better suited to assess the QPI capabilities of mixed-state ptychography. In figure~\ref{fig:muscle} we compare the reconstructed object amplitude and phase under coherent and partially coherent illumination. The coherent source is a supercontinuum laser (NKT Photonics WhiteLase Micro), which was spectrally filtered to a wavelength of $708$\,nm (bandwidth $0.6$nm) to achieve high temporal coherence. For the DMD-defined LED source we used a Costas array with $w=10\,$dpx. The DMD reconstruction used 25 modes and the laser reconstruction a single mode. For further details see~\ref{sec:recdetails}. The reconstructed single mode object image was scaled to the same pixel size as the DMD reconstruction and a constant phase offset between the two images was removed. As can be seen from figures~\ref{fig:muscle}(a) and \ref{fig:muscle}(c) the two reconstructions show differences in the transmissivity of the object, which is caused by the stain (hematoxylin and eosin) of the sample. However, the quantitative phase profiles shown in figures~\ref{fig:muscle}(b) and \ref{fig:muscle}(d) show an excellent agreement. This result affirms the reproducibility of mixed-state as compared to fully coherent ptychography.

\section{Conclusion and discussion}
In this article, we presented a technique for jointly analyzing and controlling spatially partially coherent beams, relying on mixed-state ptychography and a DMD-based light source. The suitability of this measurement method for quantitative retrieval of the MI was confirmed by comparing experimental and simulation results. Using three different types of DMD-defined source shapes in a variety of sizes, an extensive coherence study  was performed by means of the MI and its associated spatial coherence length, as well as the distribution of mixed-state eigenvalues and the resulting overall coherence. Through back-propagation to the source plane, the different source shapes could be recovered successfully, offering an additional consistency check of the method. Furthermore, we systematically studied the relationship between the source size and the recovered object spatial resolution. The results suggest that for small illumination angles the retrieved object resolution is mainly influenced by the light flux and not by the degree of coherence. However, despite requiring a longer reconstruction time, lower coherence was found not to be detrimental for spatial resolution, as suggested elsewhere~\cite{stachnik2015influence}. For imaging applications, this means that in cases where the flux and coherence quantities are linked, one has the choice between reconstruction speed (low flux, high coherence) and high spatial resolution (high flux, low coherence). In a final measurement the quantitative phase imaging ability of mixed-state ptychography was demonstrated, showing an excellent agreement with a laser-based, single-mode reconstruction.

The presented method is robust and allows the measurement of a broad class of spatially partially coherent beams with an arbitrary spatial structure. Contrary to other newly proposed coherence measurement approaches mixed-state ptychography is faster since it samples the full beam at once~\cite{divitt2014measuring, partanen2014coherence, kondakci2017coherence, naraghi2017wide}. Moreover, being a lensless technique, it requires less sophisticated optical elements as compared to LSSI and PST and the operation wavelength is not limited to the visible range~\cite{waller2012phase,camara2013optical,naraghi2017wide}. A possible future improvement is a further increase in operation speed. This could be accomplished by incorporating a priori knowledge. As an example, Schell model sources can be represented as Toeplitz matrices, which could be used as a constraint during the reconstruction process~\cite{ozaktas2002linear}. In addition, when one is only interested in characterizing the probe, the object can be precharacterized and provided as an initial guess, which significantly speeds up the convergence of the reconstruction. Together with further advancements in GPU hardware and software performance~\cite{datta2019computational}, we envision fast operation of the proposed technique, which would enable closed-loop coherence control. 

\section*{Acknowledgments} 
The authors would like to thank Bob Krijger (AMOLF) for helpful discussions regarding the operation of the DMD. This work is part of the research programme of the Netherlands Organisation for Scientific Research (NWO) and was performed at the research institute AMOLF, as well as at ARCNL, a public-private partnership of UvA, VU, NWO, and ASML. SMW and LL are partially supported by the Nederlandse Organisatie voor Wetenschappelijk Onderzoek (13934) and the European Research Council (637476).

\clearpage 
\appendix
\section*{Appendix}
\section{Setup details}
\label{sec:setupdetails}
\subsection{Measurement procedure}
In ptychography a sample is laterally scanned through a stationary beam. The step size of the scan determines the overlap between successively illuminated areas on the sample and therefore the redundancy in the ptychographic data set. The linear probe overlap $\eta$ is calculated using the FWHM of the probe $\sigma_p$  and the scan step size $s$ 

\begin{equation}
\eta = \frac{ \sigma_p - s }{\sigma_p }. 
  \label{eq:ovelap}
\end{equation}
For the spatially partially coherent illumination condition studied here the overlap was approximately $\eta=70$\,\%~\cite{bunk2008influence,burdet2015evaluation}. For the DMD-based illumination shown in figure~\ref{fig:coherence_charact} of the main article, a concentric scan grid with $260$ points and a step size of $70$\,\textmu m was used. The size variation in the three investigated source geometries caused variable overlap in the object plane. Table~\ref{tab:probe_overlap} gives an overview over the encountered probe size and overlap values for the smallest and largest source size of each shape.

\begin{figure}[t!]
\centering
{\includegraphics[width=0.95\linewidth]{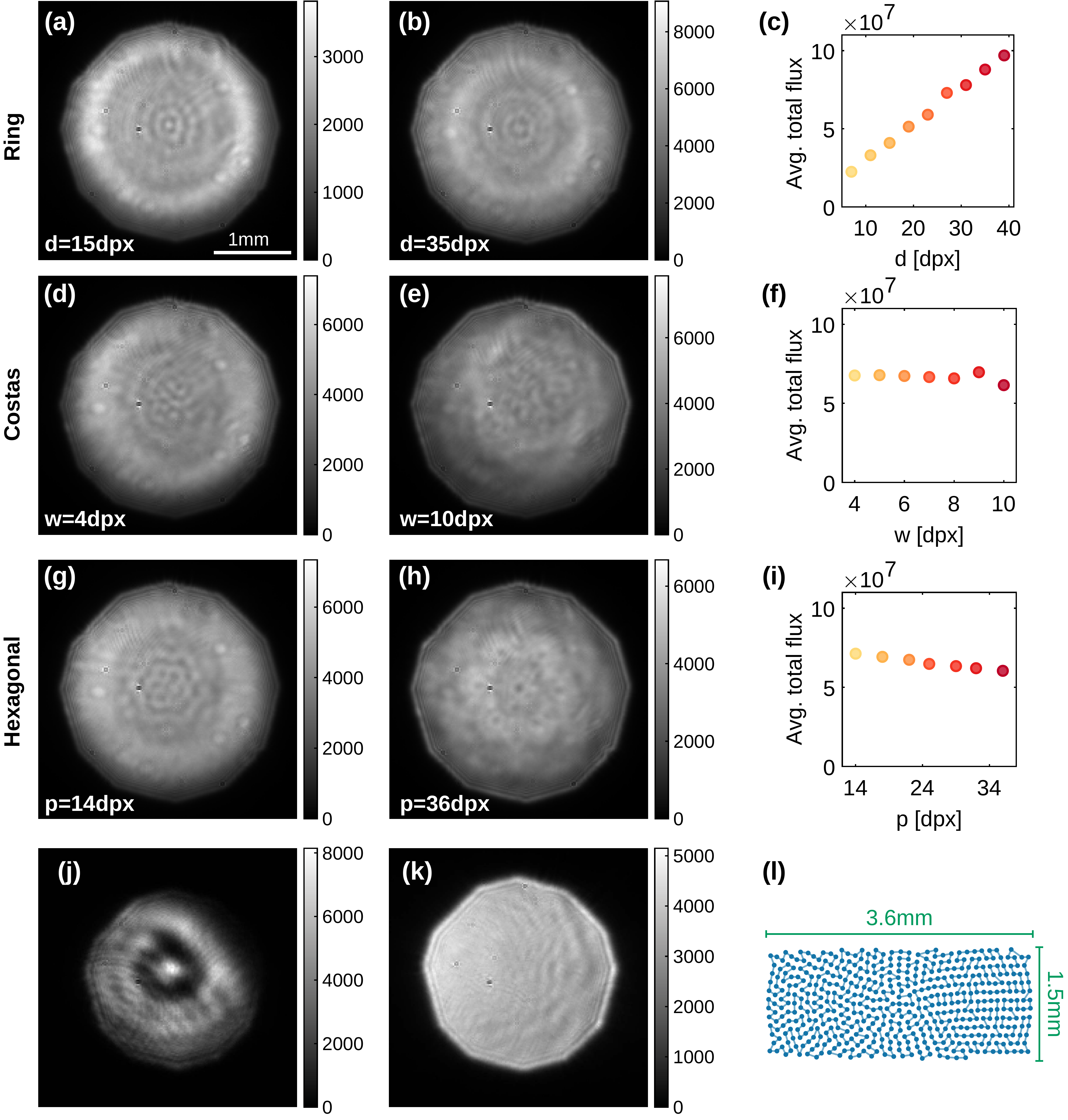}}
\caption{ Average diffraction pattern for two differently sized ring (a-b), Costas (d-e) and hexagonal (g-h) array sources. (c),(f) and (i) Average integrated photon flux as a function of the three scaling parameters $d$,$w$ and $p$. (j-k) Average diffraction pattern with the filtered supercontinuum laser (j) and DMD-defined LED (k) illumination that were used for the muscle section measurement. (l) Concentric scan grid with a rectangular FOV that was used for the muscle section measurement.}
\label{fig:S1_flux}
\end{figure}

Figure~\ref{fig:S1_flux}(a-b), \ref{fig:S1_flux}(d-e) and \ref{fig:S1_flux}(g-h) show averages over the 260 recorded diffraction patterns, each for two sizes of the ring, Costas and hexagonal sources, respectively. The exposure time was 18\,ms for the ring-shaped sources and to 19\,ms for the Costas and hexagonal array sources. To estimate the available photon flux in the different data sets we  plot the average integrated photon flux as a function of the three scaling parameters in figures~\ref{fig:S1_flux}(c),\ref{fig:S1_flux}(f) and \ref{fig:S1_flux}(i). These values were calculated by integrating over all pixels of each data set and dividing by the number of scan points.

\begin{table}[b!]
\centering
\caption{Smallest and Largest probe size $\sigma_{x,y}$ and probe overlap $\eta$ during the size sweep of the three different source shapes. $\sigma_{x,y}$ denotes the minimum of the FWHM in $x$ and $y$ direction.}
\begin{tabular}{lcccc}
\hline
Source shape & $\sigma_{x,y}$ (smallest size)  & $\sigma_{x,y}$ (largest size) & $\eta$ (smallest size) & $\eta$ (largest size) \\ \hline
Ring         & $421$\,\textmu m & $262$\,\textmu m & $83\,\%$ & $73\,\%$ \\
Costas       & $394$\,\textmu m & $394$\,\textmu m & $82\,\%$ & $82\,\%$ \\
Hexagonal    & $277$\,\textmu m & $270$\,\textmu m & $75\,\%$ & $74\,\%$ \\
\hline
\end{tabular}
  \label{tab:probe_overlap}
\end{table}

For the measurements of the histological section of a human muscle shown in figure~\ref{fig:muscle} of the article the scan grid was changed to encompass a larger field of view (FOV). Both the SC laser and the DMD measurement used a concentric grid with an FOV of $3.6\times1.5$\,mm$^2$, $496$ scan points and a step size of $95$\,\textmu m, as shown in figure~\ref{fig:S1_flux}(l). For the DMD illumination we used a Costas array source with $w=10$\,dpx. The probe had a FWHM of $440$\,\textmu m in the object plane, which resulted in a beam overlap of $78\,\%$. This higher probe beam size was achieved by moving the focusing lens further away from the object. For the laser-based illumination the probe had a FWHM of $160$\,\textmu m, which resulted in a beam overlap of $41\,\%$. In addition, for this measurement polarization filters were introduced before and after the object to rule out birefringence as a source for mixed states~\cite{ferrand2018quantitative}. In previous measurements this was not done since the object (USAF target) was assumed not to be birefringent. The exposure time was set to $9$\,ms for the laser and $180$\,ms for the DMD measurement, which resulted in  average photoelectron counts of $3.5\cdot10^7$ for the laser and $5.6\cdot10^7$ for the DMD measurement.
An average of the recorded $496$ diffraction patterns for laser and DMD-based illumination are shown in figures~\ref{fig:S1_flux}(j) and \ref{fig:S1_flux}(k). 

The CCD image acquisition and stage scan procedure is automatically controlled by an in-house developed .NET software. Background signals were removed by subtracting a CCD image, which was acquired with a blocked DMD beam. 

\subsection{DMD control}
\label{sec:DMD}
The Digital Micromirror Device (DMD) consists of an array of almost a million micromirrors, which each can be set to an on or off state. The DMD is connected via HDMI to a computer as a second screen. The pattern is controlled using a Python GUI we implemented using wxPython and code from Ref.~\cite{popoff2017}. To avoid pixel interpolation the Python script creates a window with the same resolution as the DMD ($1280\times720$). Further DMD and LED settings were controlled with the DLP Display and Light Control EVM GUI tool by Texas Instruments. At low exposure times flickering can occur due to a finite dark time of the DMD micromirrors, unless the DMD and CCD are synchronized. We attribute the ensemble coherence effects mentioned in the main text to be due to switching between the dark and bright states of the DMD. 

\begin{figure}[t!]
\centering
{\includegraphics[width=0.55\linewidth]{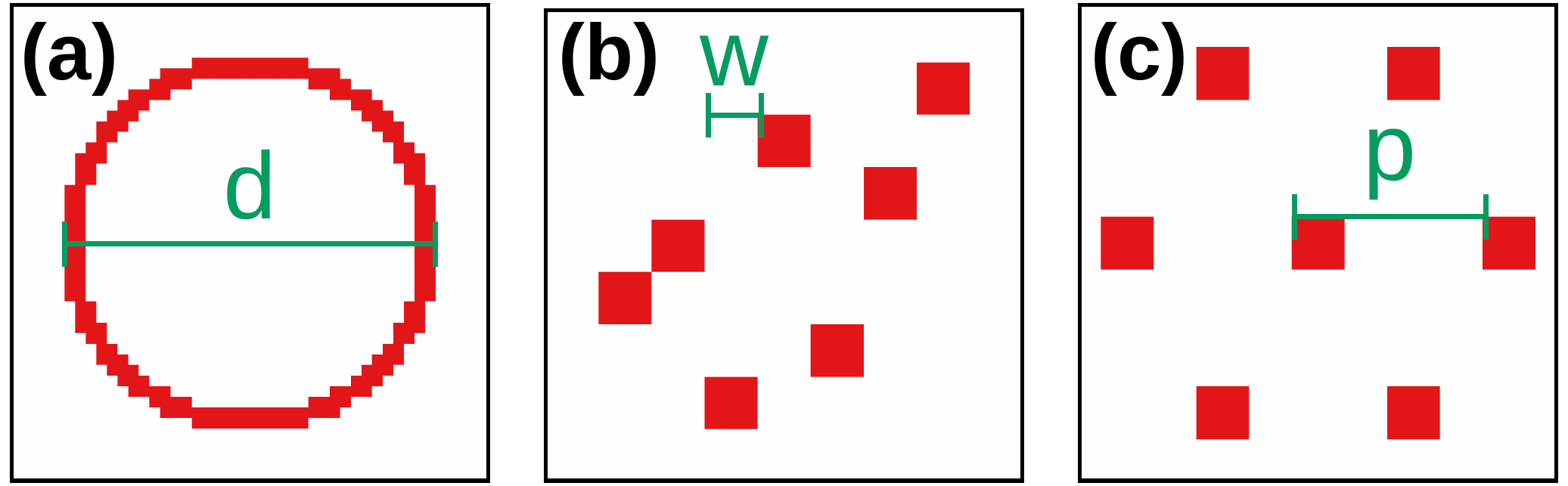}}
\caption{Examples of DMD images used to create the ring, Costas and hexagonal array sources. Shown in false color.  The indicated parameters ($d,w,p$) were used to scale the three source shapes.}
\label{fig:S3_DMD}
\end{figure}

Figures~\ref{fig:S3_DMD}(a) to \ref{fig:S3_DMD}(c) show cropped false color versions of 3 different DMD images used to create the ring, Costas and hexagonal array sources. The indicated parameters ($d,w,p$) were used to scale the three source shapes. The ring-shaped DMD images had a width, which remained $2$\,dpx and an outer diameter $d$, which was varied between $7\,$dpx and $39\,$dpx in steps of $4\,$dpx. For the Costas array, the source size was varied by increasing  the distance $w$ from $4\,$dpx to $10\,$dpx in steps of $1\,$dpx, while the edge length of each of the $7$ square elements stayed $5\,$dpx. The size parameter $p$ of the hexagonal array was chosen such that the largest distances in the Costas and hexagonal arrays approximately matched each other. For visual clarity, figures~\ref{fig:coherence_charact}(a) to \ref{fig:coherence_charact}(i) in the main article show reconstruction results of every second measured and reconstructed source size, while the respective inset plots show extracted metrics for the complete size sweep.

\section{Reconstruction details}
\label{sec:recdetails}
The mixed-state ptychography algorithm was implemented in MATLAB. The feedback parameter $\beta$ in Eqs.~\eref{eq:Pupdate} and~\eref{eq:Oupdate} in the main text is set to $0.25$. As an initial guess of the probe reconstruction a binary circle of $300$\,\textmu m diameter is used. 

Real-space sampling in the object plane is given by
\begin{equation}
\Delta r = \frac{\lambda z_o}{N \Delta q},
  \label{eq:pixelsize}
\end{equation}
where $\lambda$ is the wavelength, $z_o$ the object-detector distance, $N$ the number of detector pixels, and $\Delta q$ the detector pixel size.
In addition to that, $\Delta q$ also influences the probe field of view (pFOV), which is given by
\begin{equation}
\mathrm{pFOV} = \frac{\lambda z_o }{\Delta q}.
  \label{eq:pFOV}
\end{equation}
We binned the detector images by a factor of $4$, which leads to an effective detector pixel size of $\Delta q = 4.54\,$\textmu m$\times 4=18.16\,$\textmu m, while significantly reducing the reconstruction time. In case of our DMD-based measurements the binning resulted in a object plane sampling size of $\Delta r=3.6\,$\textmu m and a probe FOV of $\mathrm{pFOV}=1.31\,$mm, which was still large enough to fully capture the probe beam in the object plane. 

\begin{figure}[t!]
\centering
{\includegraphics[width=1\linewidth]{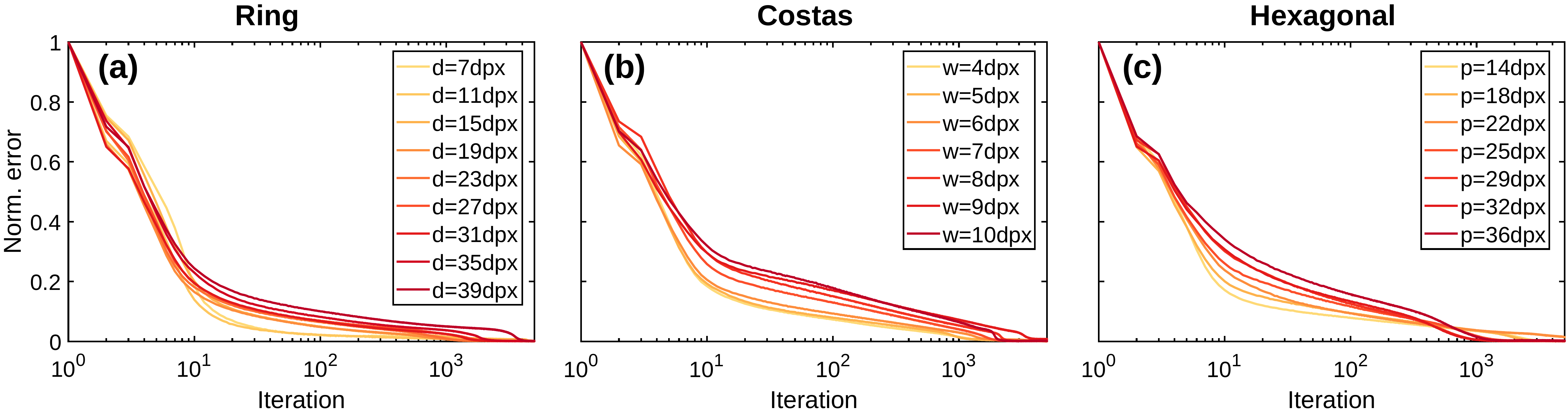}}
\caption{Convergence of the normalized error during the reconstruction of data sets using ring, Costas and hexagonal array sources of different sizes. The x-axis is shown on a logarithmic scale.}
\label{fig:convergence}
\end{figure}

Each reconstruction for the DMD based source variations was run for $5000$ iterations. In order to monitor the convergence during reconstruction, an error metric is calculated at every iteration as the difference between the measured and estimated diffraction patterns
\begin{equation}
\epsilon_{n}=\sum_{j}\left|I_{\mathrm{meas},j}\left(\boldsymbol{q}\right)-\sum_{m}\left|\tilde{\psi}_{m,j}^{n}\left(\boldsymbol{q}\right)\right|^{2}\right|.
  \label{eq:error}
\end{equation}
Figure~\ref{fig:convergence} shows the evolution of the error curves normalized to their maximum  for reconstruction of the three source shape size sweeps. One can see that for more incoherent cases  the reconstructions tend to converge slower. 
 
\begin{figure}[t!]
\centering
{\includegraphics[width=0.6\linewidth]{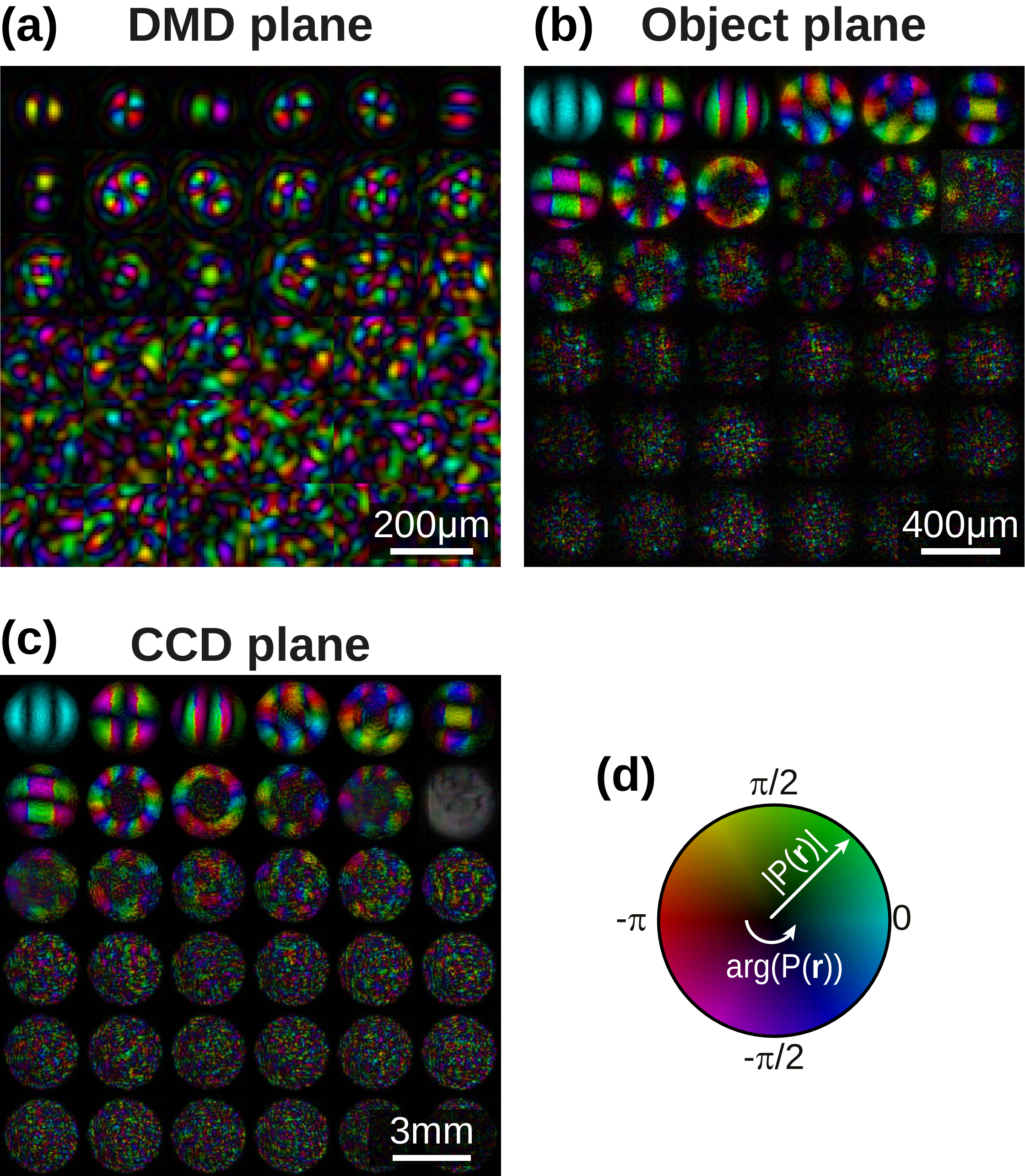}}
\caption{ (a-c) Reconstructed 36 coherent modes for a ring-shaped source with $d=15$\,dpx in the DMD, object and CCD plane, respectively. The phase curvature of the probe modes in panels (b) and (c) was removed. (d) A circular representation of the color map used to jointly represent amplitude and phase of the mixed-modes.}
\label{fig:S4_modes}
\end{figure}

During the reconstruction in intervals of $10$ iterations the SVD-based orthogonalization of the probe is carried out~\cite{loetgering2017data}. The number of modes $L$ was kept at $36$ for most of the reconstructions. In Figures~\ref{fig:S4_modes}(a) to \ref{fig:S4_modes}(c) we show all 36 reconstructed modes in the DMD, object and CCD plane, respectively for the data set with a ring-shaped source with $d=15$\,dpx . The quadratic phase curvature of the modes shown in figures~\ref{fig:S4_modes}(b) and \ref{fig:S4_modes}(c) was removed through multiplication with the conjugate phase of the primary coherent mode. For the reconstructions with ring-shaped illumination with a diameter $d=[27,31,35,39]$\,dpx the mode number was increased to $49$ in order to take the lower overall coherence of those data sets into account. The MATLAB reconstruction code was accelerated using NVIDIA CUDA. A reconstruction with 36 modes running for 5000 iterations took around 30 minutes on a workstation with a NVIDIA Geforce RTX 2080 Ti GPU.

For a quantitative resolution assessment of the reconstructed complex object function we use the Fourier ring correlation (FRC) approach. The FRC results shown in figures~\ref{fig:coherence_charact}(g)-\ref{fig:coherence_charact}(i) of the article were calculated using
\begin{equation}
{\mathrm{FRC}} (q) = \frac{{\mathop {\sum}\limits_{ q'\in R_q} \tilde{O}^\ast_1(q') \cdot \tilde{O}_2(q' ) }}{{\sqrt {\mathop {\sum}\limits_{q'\in R_q} \left|\tilde{O}_1(q')\right|^2  \cdot \mathop {\sum}\limits_{q'\in R_q } \left|\tilde{O}_2(q' )\right|^2} }},
 \label{eq:FRC}
\end{equation}
where $q$ denotes a spatial frequency magnitude, $R_q$ defines a set of Fourier space pixels in a ring with radius $q$, $\tilde{O}_{1}$ and $\tilde{O}_{1}$ are the Fourier transformed object functions from two independent data sets. These two independent object reconstructions were obtained by performing 2000 mPIE iterations on two halves of each data set (130 images). Before the FRC is calculated the two images are spatially aligned using subpixel registration, the phase offsets are synchronized and a Hann window is applied, which removes reconstruction artifacts outside of the FOV. 

\section*{References}
\bibliographystyle{iopart-num}
\bibliography{references}

\end{document}